\documentclass[11pt]{article}
\usepackage[margin=1in]{geometry}
\usepackage{graphicx}

\usepackage{comment}
\usepackage{lineno}

\def\beq{\begin{equation}}
\def\eeq#1{\label{#1}\end{equation}}
\def\eeqn{\end{equation}}

\def\beqa{\begin{eqnarray}}
\def\eeqa#1{\label{#1}\end{eqnarray}}
\def\eeqan{\end{eqnarray}}

\let\bar=\overbar

\def\Dslash{\not{\hbox{\kern-4pt $D$}}}
\def\dslash{\not{\hbox{\kern-2pt $\del$}}}

\def\msb{{\bar{\ssstyle M \kern -1pt S}}}

\def\Title#1{\begin{center} {\Large {\bf #1} } \end{center}}
\def\Author#1{\begin{center} {\normalsize {\sc #1} } \end{center}}
\def\Institution#1{\begin{center} {\normalsize {\it #1} } \end{center}}
\def\Abstract#1{\noindent {\normalsize {\bf Abstract:} {\normalfont #1}}}
\def\Conference{\vspace{4mm}\begin{raggedright} {\normalsize {\it Talk presented at the 2019 Meeting of the Division of Particles and Fields of the American Physical Society (DPF2019), July 29--August 2, 2019, Northeastern University, Boston, C1907293.} } \end{raggedright}\vspace{4mm}}

\begin{document}

%
%

\Title{Search for direct stop pair production with the ATLAS detector}

\Author{Keisuke Yoshihara}

\Institution{Department of Physics and Astronomy\\ University of Pennsylvania, Philadelphia, PA, USA}

\Abstract{Supersymmetry, which extends the Standard Model (SM) by introducing supersymmetric partners for the SM particles, can provide an elegant solution to the hierarchy problem. One of the most important parameters in supersymmetry is the mass of the supersymmetric partner to the top quark, referred to as stop. In the absence of the stop signature in the previous searches at the LHC, compressed stop searches in which the mass of the stop is close to the mass of the lightest neutralino become more important. In scenarios with compressed mass spectra, the momentum transfer to the decay products of the stop may be small, leading to decay products with transverse momenta of only a few GeV. Hence, the identification and reconstruction of those ``soft'' objects can play a crucial role in the searches. In this article, recent results from the compressed stop searches in $pp$ collisions at $\sqrt{s}=$13 TeV with the ATLAS detector and new developments for soft $b$-hadron tagging techniques are presented.}

\Conference

%
%

\section{Introduction}

\begin{figure}[htb]
\centering
\includegraphics[width=.75\textwidth]{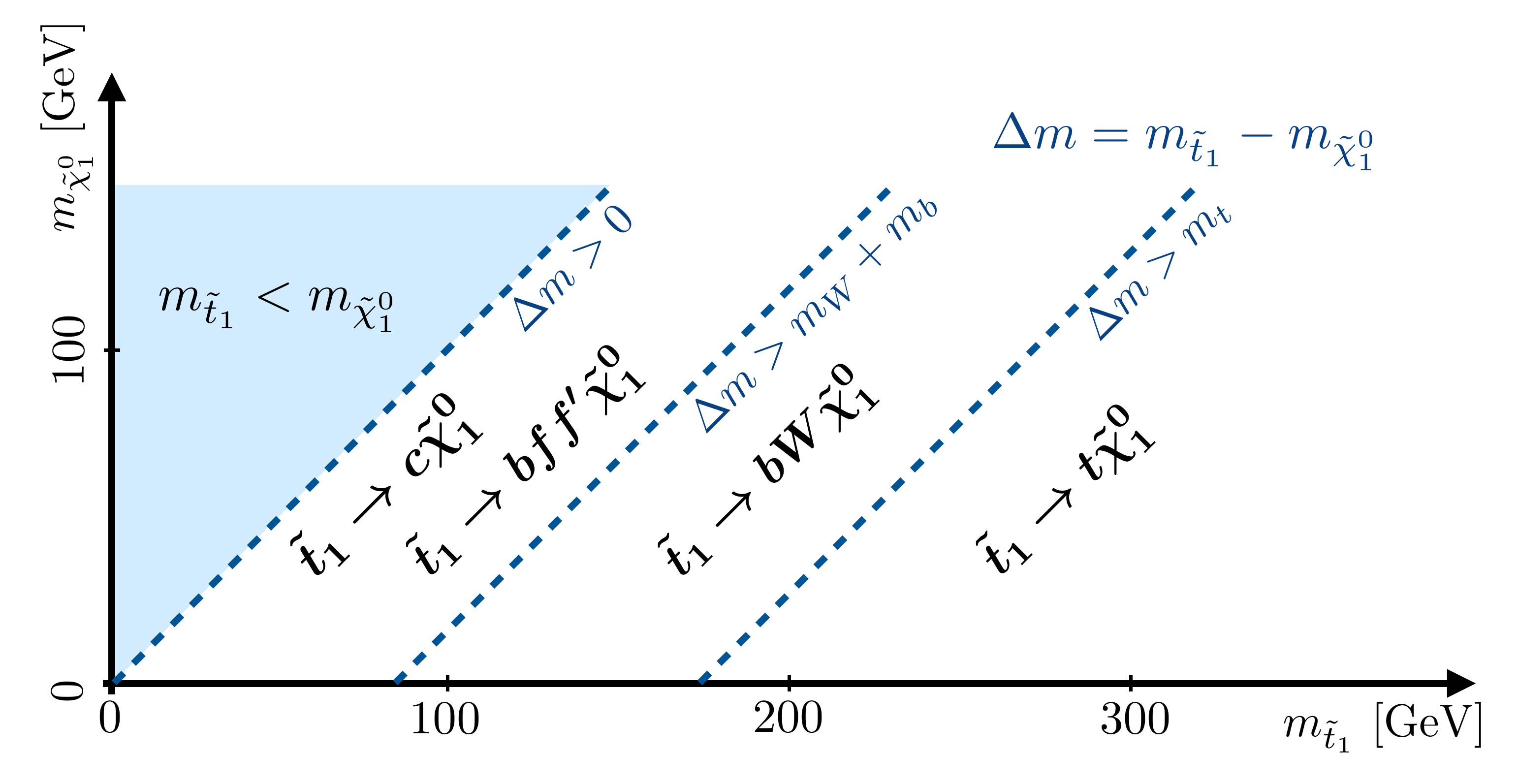}
\caption{Illustration of the preferred stop decay modes in the
plane $\tilde{t}_{1}$ - $\tilde{\chi}^{0}_{1}$ mass plane. The neutralino is assumed to be the lightest supersymmetric particle~\cite{paper:Stop1L2017}.}
\label{fig:tN_phase_space}
\end{figure}

\noindent Searches for a light supersymmetric partner of the top quark, denoted as the top squark or stop, are of particular interest after the discovery of the Higgs boson at the LHC\cite{LHC}.
Top squarks may largely cancel divergent loop corrections to the Higgs-boson mass, and thus, supersymmetry may provide an elegant solution to the hierarchy problem. 
The superpartners of the left- and right-handed top quarks, $\tilde{t}_{L}$ and $\tilde{t}_{R}$, mix and form two mass eigenstates, $\tilde{t}_{1}$ and $\tilde{t}_{2}$, where $\tilde{t}_{1}$ is lighter of the two.
A generic R-parity-conserving minimal supersymmetric extension of the SM (MSSM) predicts pair production of SUSY particles and the existence of a stable lightest supersymmetric particle (LSP). 
The mass eigenstates from the linear superposition of charged and neutral SUSY partners of the Higgs and electroweak gauge bosons (Higgsinos, winos and binos) are
called charginos $\tilde{\chi}^{\pm}$ and neutralinos $\tilde{\chi}^{0}$. In targeted SUSY models, stops are pair produced and the lightest neutralino, $\tilde{\chi}^{0}_{1}$, is assumed to be the LSP and a potential dark matter candidate. 
\\\\
Figure~\ref{fig:tN_phase_space} illustrates the kinematically allowed phase space of the stop decay, which is defined by the mass-splitting $\Delta m = m(\tilde{t}_{1})-m(\tilde{\chi}^{0}_{1})$. 
In the region where $\Delta m$ is less than the top-quark mass but larger than the sum of the b-quark and W-boson masses, the stop undergoes a three-body decay ($bW\tilde{\chi}^{0}_{1}$). 
Such decays are favored unless the mass difference is smaller than the sum of masses of the b-quark and W-boson, in which case the decay would proceed via a four-body process ($bff'\tilde{\chi}^{0}_{1}$). In the absence of the stop signature in previous searches~\cite{paper:Stop1L2017}, more complex stop scenarios have been paid more attention such as a compressed stop signature where the mass of the LSP is close to the mass of the stop. In this article, (1) a novel analysis targeting the stop three-body decay scenario with a machine learning technique, and (2) new developments of soft $b$-hadron tagging techniques targeting the stop four-body decay scenario are presented. The analyses are based on 139fb$^{-1}$ of $\sqrt{s}=13$ TeV proton-proton collision data recorded with the ATLAS detector\cite{ATLAS} at the LHC from 2015 to 2018.

\section{Three-body search using a machine-learning technique}

\noindent The three-body signal scenario, $\tilde{t}_{1} \rightarrow bW\tilde{\chi}^{0}_{1}$, is a simplified model in which the masses of all sparticles are set to high values except for the two sparticles involved in the decay chain of interest. 

\subsection*{Event Selection}

\noindent All events were selected with triggers based on the presence of large missing transverse energy $E_{T}^{miss}$, which are fully efficient for events passing an offline-reconstructed $E_{T}^{miss}$ $>$ 230 GeV requirement~\cite{CONF:Stop1L2019}. All events are required to have exactly one charged lepton ($e$ or $\mu$) with large transverse momentum $p_T$ satisfying strict identification criteria~\cite{PUB:ElectronID, PUB:MuonID}, no additional leptons and four or more jets, at least one of which must be a $b$-tagged jet. In order to reject multijet events, a requirement is imposed on the azimuthal angles between the leading and sub-leading jets and $E_{T}^{miss}$, $\Delta\phi(j_{1,2}, E_{T}^{miss})$. The $W$+jets and semi-leptonic $t\bar{t}$ backgrounds are largely suppressed by requiring the transverse mass, $m_T$, to be well above the W-boson mass. Finally, for events with hadronic $\tau$ candidates, the requirement $m_{T}^{\tau}$ $>$ 80 GeV is applied, where $m_T^{\tau}$ is a variant of the variable $m_{T2}$. The $m_{T2}$ variable is a generalization of the transverse mass applied to signatures where two particles are not directly detected, and specifically targets events where a W-boson decays via a hadronically decaying $\tau$ lepton. 

\subsection*{SR selection using ML classifier}

\noindent  The dominant background process after the preselection above is the dileptonic $t\bar{t}$ process. In order to better discriminate the signal from the large amount of background, the signal region (SR) selection is optimized by employing a machine learning (ML) approach. The ML classifier maximizes the search sensitivity by learning the difference in the event topology between the signal and the $t\bar{t}$ background and by extracting correlations amongst a set of kinematic distributions being considered as input for the training. The size of the training sample is a crucial aspect for the performance of any ML method. For the signal, since generating the full simulation samples with adequate sample sizes is computationally expensive, events without detector simulation were used for the training to boost the statistics by two orders of magnitude. The generated events were then smeared using a dedicated procedure to emulate the effects of detector simulation and reconstruction. For the SM background processes, fully simulated and reconstructed events were used, as they contain a sufficient number of events for training the classifier.
\\\\
According to the kinematics of the signal model, the jet multiplicity in the final states may vary significantly in the population of signal events. In order to deal with the variable length signal jet collection, the first step of the ML architecture employs a recurrent neural network (RNN). A key benefit of RNNs is the ability to extract information from sequences of arbitrary length. In the second step, the output vector of the jet-based RNN is passed to a shallow neural network (NN). At this stage, additional discriminating variables are also considered and passed to the NN as input variables to improve the discriminating power of the ML classifier. The SR is defined by a stringent requirement on the classifier output score, NN$_{bWN}>0.90$. While for the exclusion, the SR is expanded to lower bins down to 0.65, exploiting the shape of the NN$_{bWN}$ distribution. For the three lowest bins in the shape-fit configuration, an additional requirement of $m_T$ $>$ 150 GeV is applied to suppress potential contamination from semi-leptonic $t\bar{t}$ events.

\subsection*{Background estimate}

\noindent The dominant dileptonic $t\bar{t}$ background is normalized in a high-purity control region (CR) defined by relaxing the selection requirement on the output score of the ML classifier to 0.40--0.60. In addition, the requirement on $m_T$ is tightened to $m_T$ $>$150 GeV to reduce the semi-leptonic $t\bar{t}$ contamination. The background estimate is then tested using a validation region (VR), which is disjoint from the CR and SR. The VR is defined by sliding the output score window to 0.60--0.65. In addition, the $m_T$ $>$ 150 GeV requirement is kept to suppress the semi-leptonic $t\bar{t}$ contamination. 

\subsection*{Result}

\begin{figure}[!htb]
\centering
\includegraphics[width=.37\textwidth]{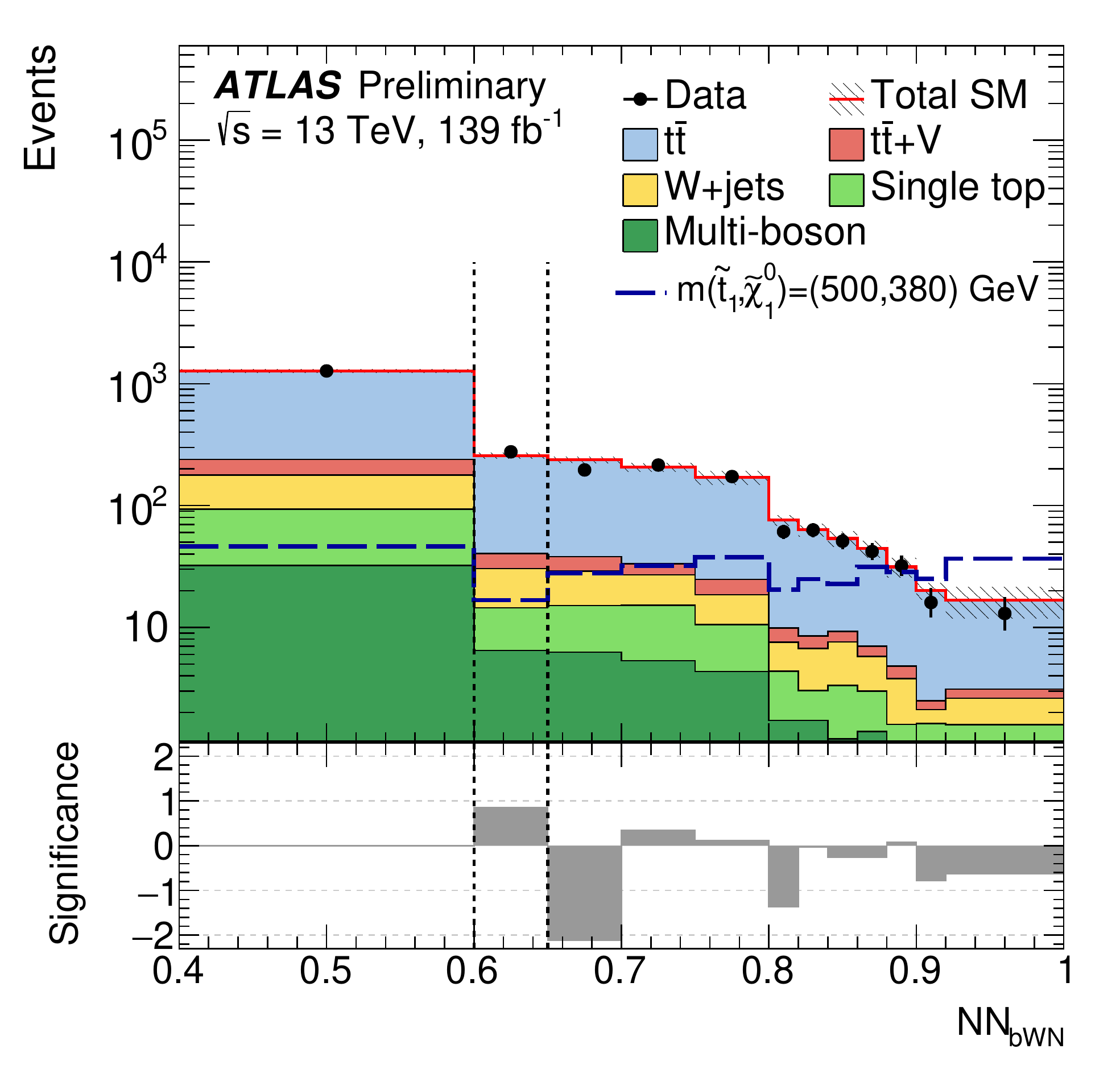}
\includegraphics[width=.50\textwidth]{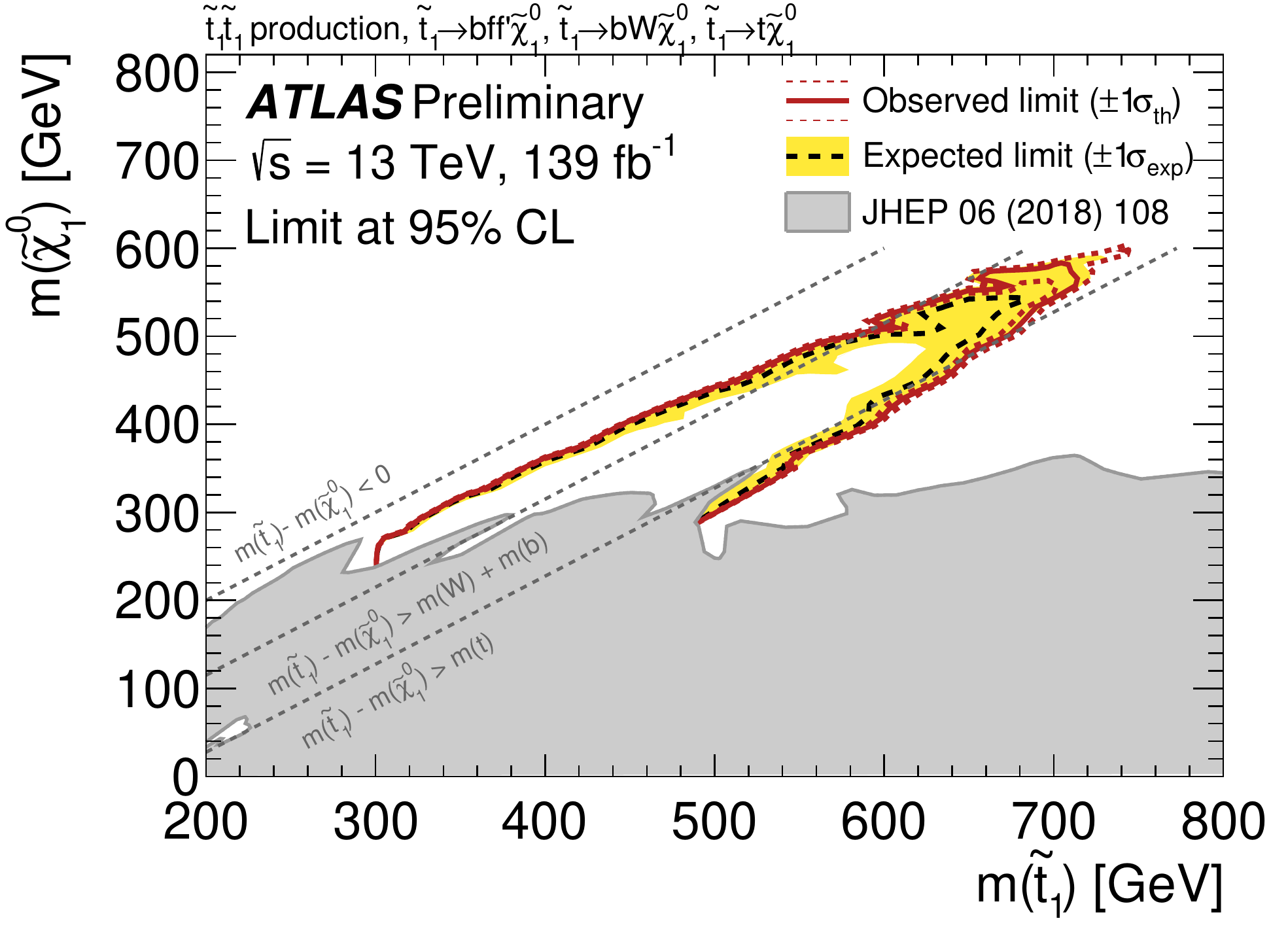}
\caption{Left: Distribution of output score of the machine learning classifier, NN$_{bWN}$. From left to right, the bins correspond to the CR, VR and the multiple bins of the SR. The boundaries of the regions are shown as vertical dashed black lines. The hatched area around the total SM background prediction includes all uncertainties. The bottom panel shows the significance $Z$ of the observed data given the predicted SM background. Right: Expected (black dashed) and observed (red solid) 95\% excluded regions in the plane of m($\tilde{\chi}^{0}_{1}$) versus m($\tilde{t}_{1}$) for direct stop pair production. The grey shaded area denotes the previously excluded regions from Ref.~\cite{paper:Stop1L2017}. Figures are taken from Ref.~\cite{CONF:Stop1L2019}.}
\label{fig:bWN_result}
\end{figure}

\noindent The likelihood fit is performed to see the agreement between observed events and the SM background prediction. The number of observed events and the predicted number of SM background events in the ML classifier distribution are shown in Figure~\ref{fig:bWN_result} (Left). The bins correspond to the CR, VR, and SR. No significant data excess is found in the SR. The observed data and the SM background prediction are found to be in good agreement. The bottom panel shows the significance of the observed data given the predicted SM background. 

\noindent Figure~\ref{fig:bWN_result} (Right) also shows the expected and observed exclusion contour as a function of stop and neutralino mass in the targeted three-body decay model. In addition, exclusion limits for signal points in the adjacent two-body and four-body regimes have been determined. The results improve upon previous exclusion limits by excluding the stop mass region up to 720 GeV for a neutralino of about 580 GeV under the assumption of 100\% {\it BR}($\tilde{t}_{1} \rightarrow bW\tilde{\chi}^{0}_{1}$).

\section{Soft $b$-hadron identification for stop four-body decay}

\begin{figure}[htb]
\centering
\includegraphics[width=.40\textwidth]{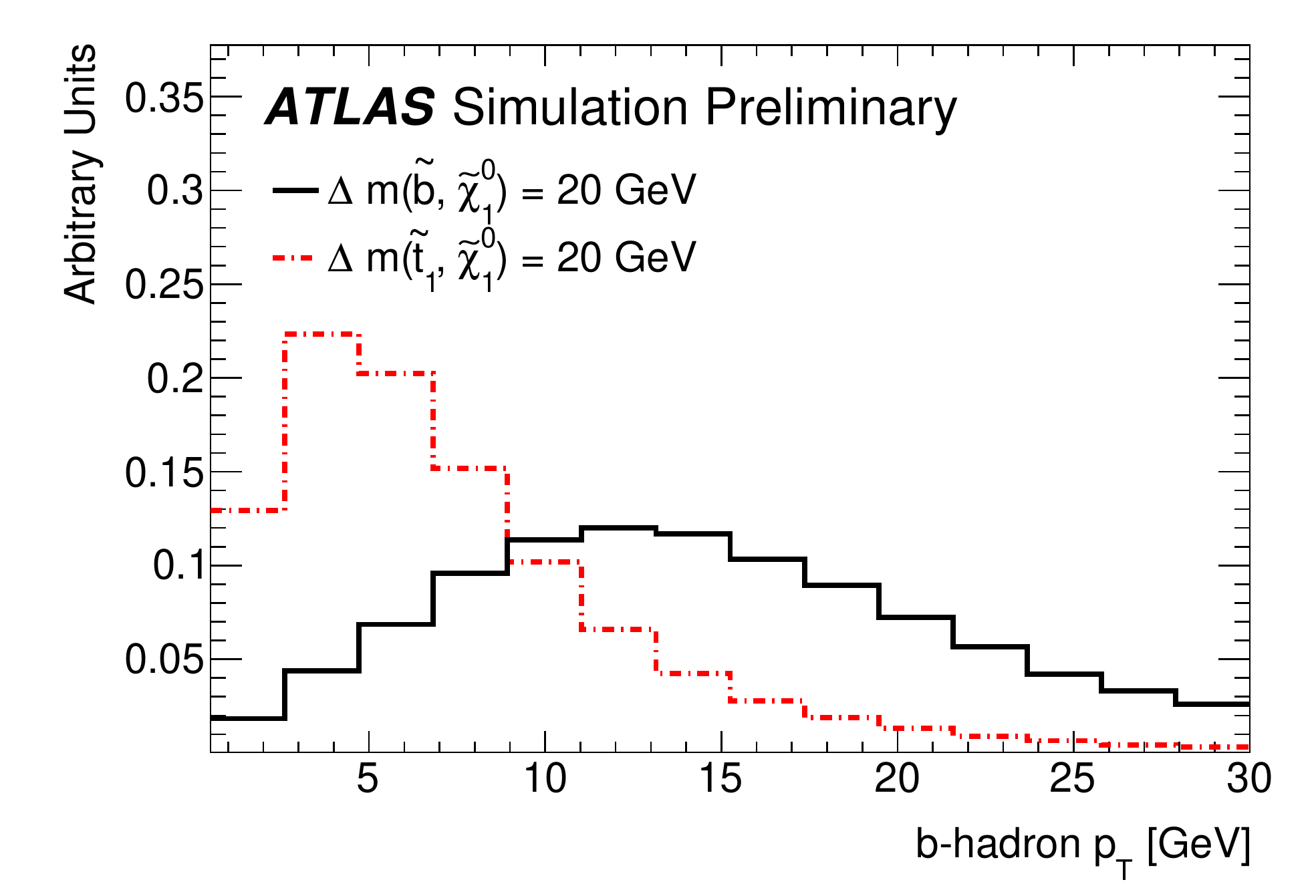}
\includegraphics[width=.40\textwidth]{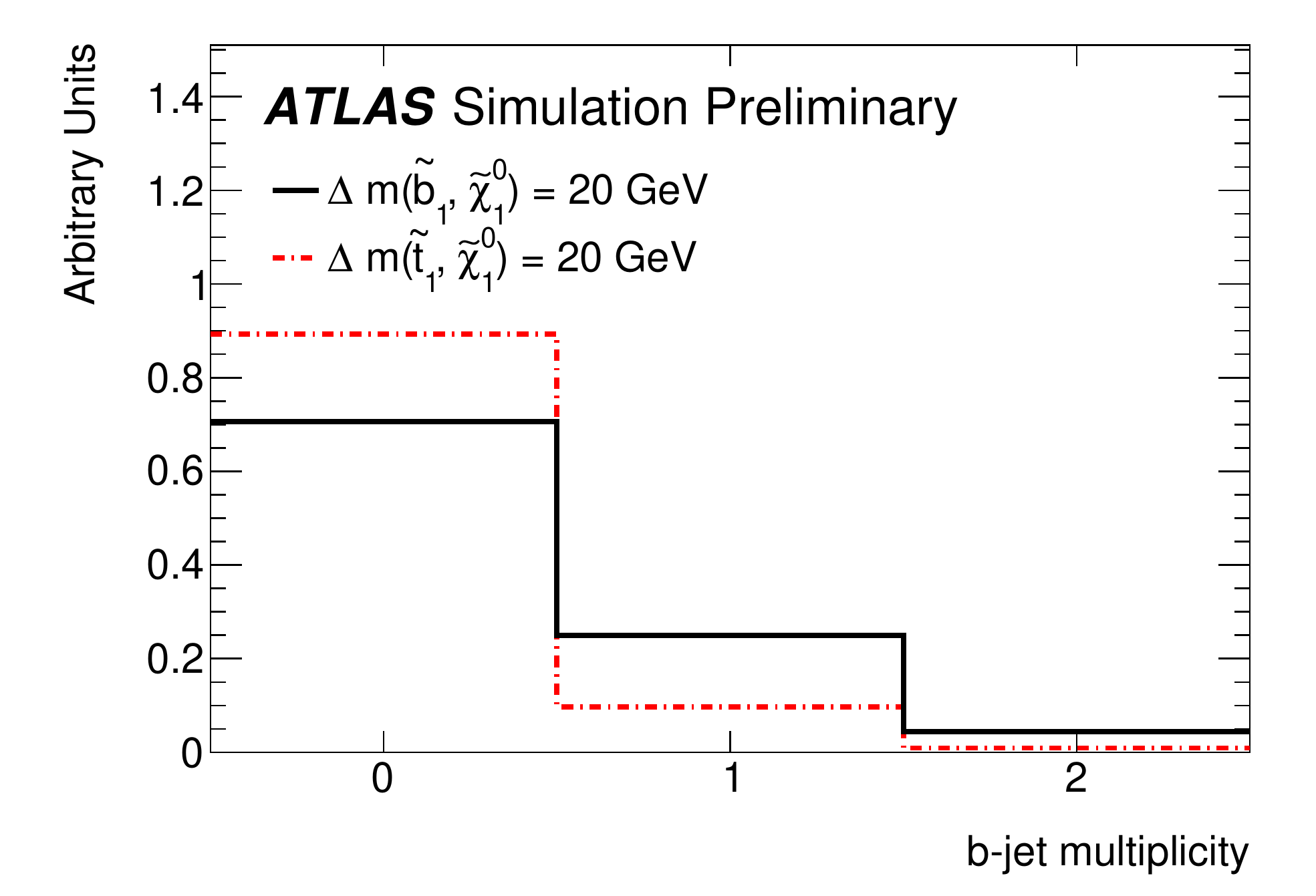}
\caption{Left: $p_T$ spectrum of $b$-hadrons produced in either $\tilde{t}_{1}\rightarrow bff' \tilde{\chi}^{0}_{1}$ or $\tilde{b}_{1}\rightarrow b \tilde{\chi}^{0}_{1}$ decay. Right: The number of $b$-tagged jets with $p_T$ $>$ 20 GeV in the ATLAS standard MV2c10 algorithm at 77\% efficiency working point~\cite{CONF:Softb2019}.}
\label{fig:softb_spectrum}
\end{figure}

\noindent In compressed SUSY scenarios such as the stop four-body decay, the stop decay products, in particular, $b$-hadrons, are expected to have very low momentum transfer as shown in Figure~\ref{fig:softb_spectrum}. These `soft' $b$-hadrons are often neither reconstructed nor identified in the detector. The identification of $b$-hadrons is crucial for reducing the enormous $W/Z$+jets background in searches for compressed stop decays. However, the ATLAS standard $b$-tagging algorithm~\cite{paper:Btagging, PUB:Btagging}, which is optimized for high-$p_T$ $b$-hadrons covering the $b$-hadron $p_T$ down to 20 GeV, may not be sufficient. A new algorithm~\cite{CONF:Softb2019} has been developed to identify soft $b$-hadrons below $p_T = 20$ GeV to help reduce these backgrounds.

\subsection*{Identification algorithm}

\noindent The identification of soft $b$-hadrons is based on the presence of secondary vertices reconstructed from tracks. Given that there are typically $O(1000)$ tracks in an event, track combinatorics is enormous, leading to a long CPU time in the vertex finding. In order to minimize the processing time, the algorithm starts with tracks that appear to be displaced from the primary $pp$ collision point based on selection criteria such as significance of track impact parameters,  $z_0/\sigma{(z_0)} > 0.5$ and $d_0/\sigma{(d_0)} > 1.7$, which help reduce the number of tracks considered for the vertexing. The number of tracks are further reduced by imposing a requirement such that tracks are well separated ($\Delta R (j, trk) < 0.4$) from the axis of a calorimeter jet with $p_T$ $>$ 30 GeV, ensuring the orthogonality to the standard $b$-tagging algorithm. The two-track seed vertices are then built from these selected tracks. All possible two-track combinations are tested, and the seeds are retained if they satisfy cos$\theta$ $>$ 0.7 where $\theta$ is the angle between the vector $\vec{r}$ pointing from the primary vertex to the secondary vertex and the 3-vector $\vec{p}^{~vtx}$ obtained from the vectorial sum of the track momenta. Starting from the selected two-track seed vertices, n-track vertices are formed ($n > 2$). At this stage, potential ambiguities arising from track association to multiple vertices are resolved. Finally additional vertex selection criteria are applied such as an invariant mass greater than 600 MeV to enhance vertices from $b$-hadron decays. 

\subsection*{Performance of the tagger}

\begin{figure}[htb]
\centering
\includegraphics[width=.40\textwidth]{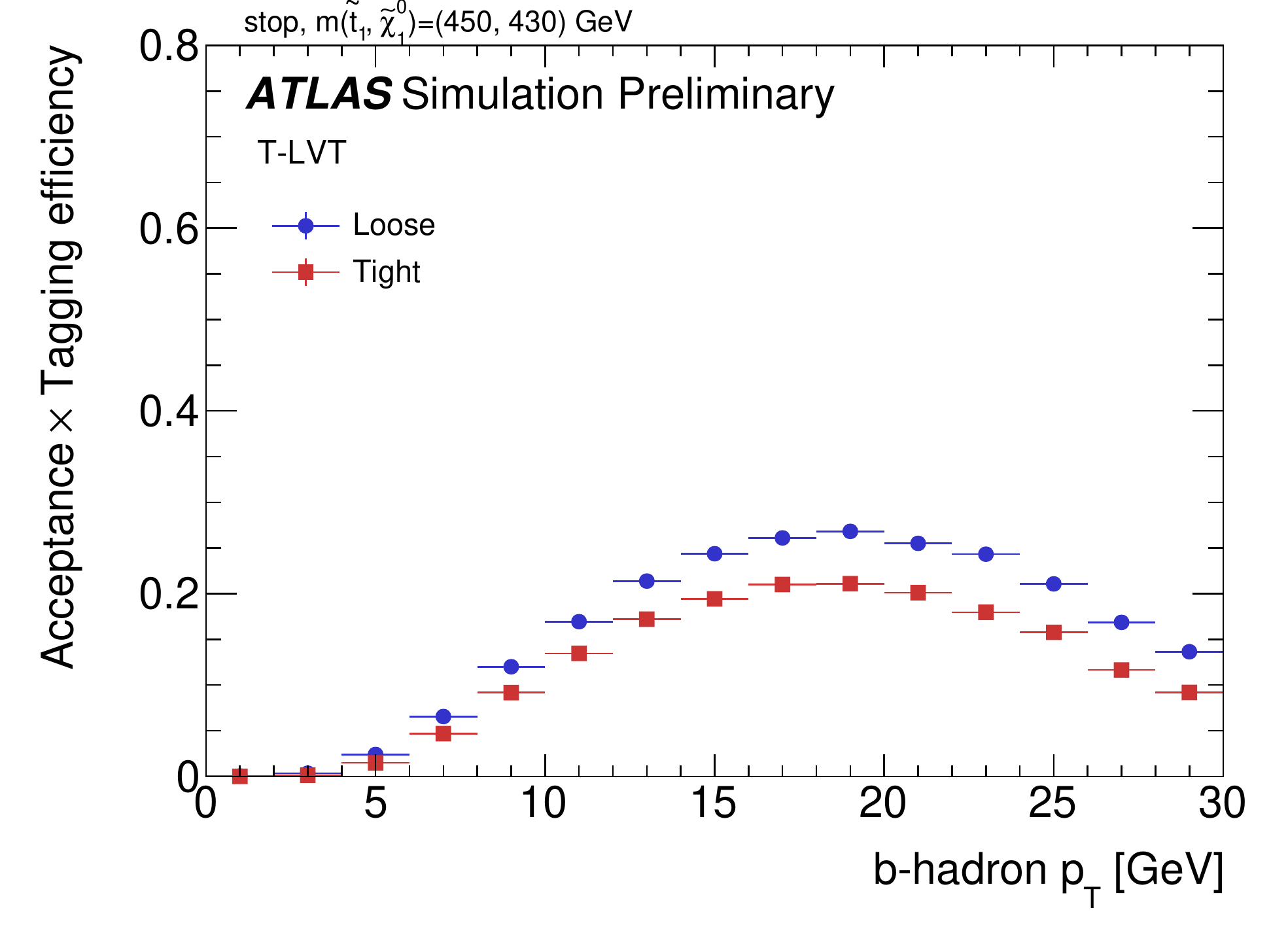}
\includegraphics[width=.40\textwidth]{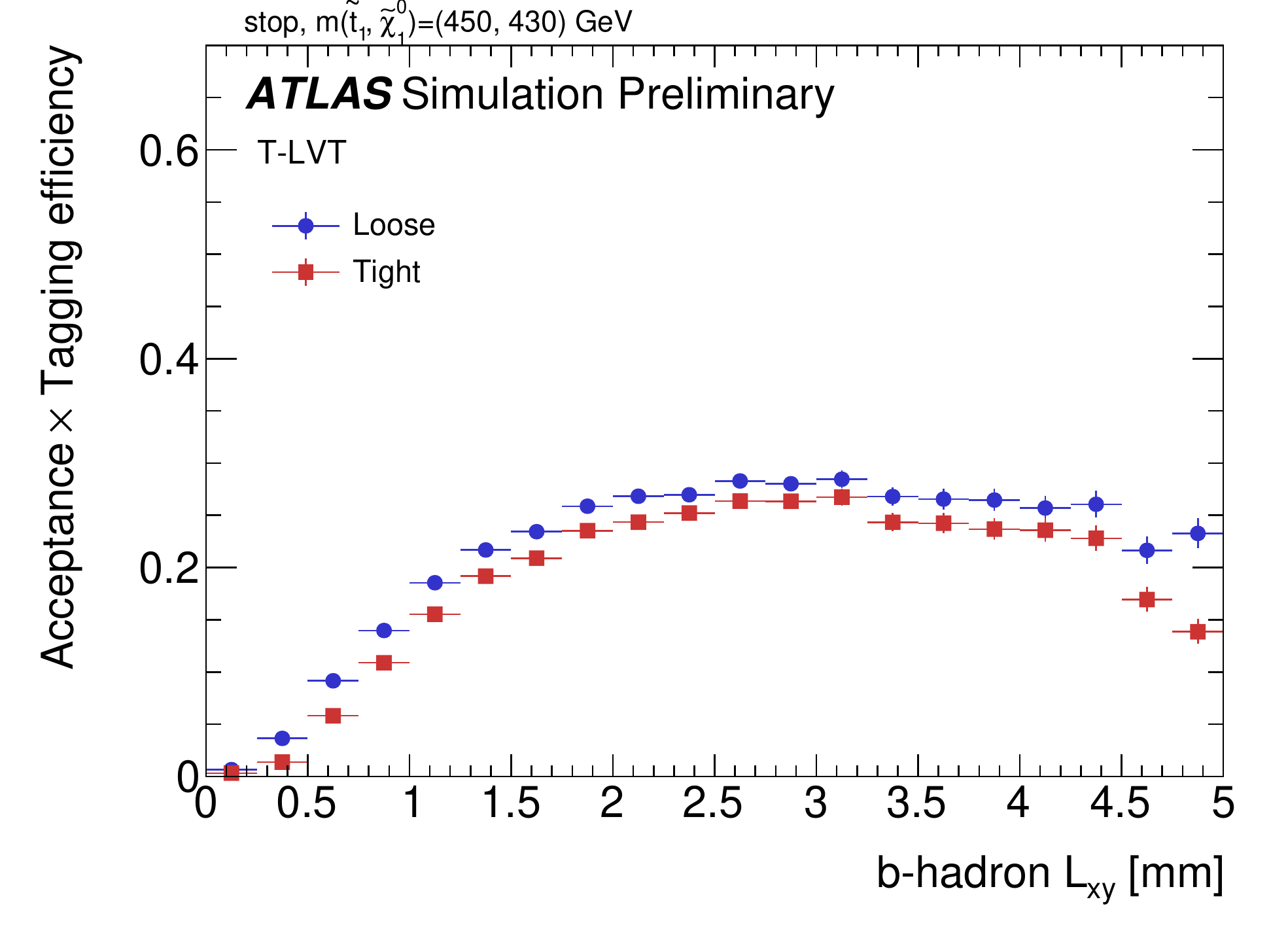}
\caption{$b$-hadron acceptance times tagging efficiency as a function of (left) the $b$-hadron $p_T$, and (right) the distance in the transverse plane from the primary vertex to the secondary vertex, $L_{xy}$~\cite{CONF:Softb2019}.}
\label{fig:softb_efficiency}
\end{figure}

\noindent The performance of the developed tagger is evaluated using simulated events. In order to measure efficiencies and fake-rates on equal footing, a common definition for the matching of a vertex to generator-level particles is defined. If a `truth' $b$-hadron is found within a cone of size $\Delta R < 0.3$ from the secondary vertex angular direction, the vertex is considered as a truth-matched $b$-hadron. The $c$-hadron matching is further tested in the same manner if the vertex does not match to a $b$-hadron. If the vertex does not fall into either of these categories, the vertex is considered as a fake vertex. The acceptance times tagging efficiency is evaluated as a function of $p_T$ or $L_{xy}$ for two working points (WP), {\it Tight} and {\it Loose}~\cite{CONF:Softb2019} as shown in Figure~\ref{fig:softb_efficiency}. For $b$-hadrons with $p_T$ between 5 GeV and 15 GeV, the fake rate is found to be $8\times10^{-3}$ and $2.7\times10^{-2}$ for {\it Tight} and {\it Loose}, respectively. For the $b$-tagging efficiency, all $b$-hadrons in simulated stop four-body signal events are included in the denominator of the efficiency calculation. While for the fake rate, all signal events are included in the denominator regardless the presence of $b$-hadrons in the event. 
%

\begin{figure}[htb]
\centering
\includegraphics[width=.40\textwidth]{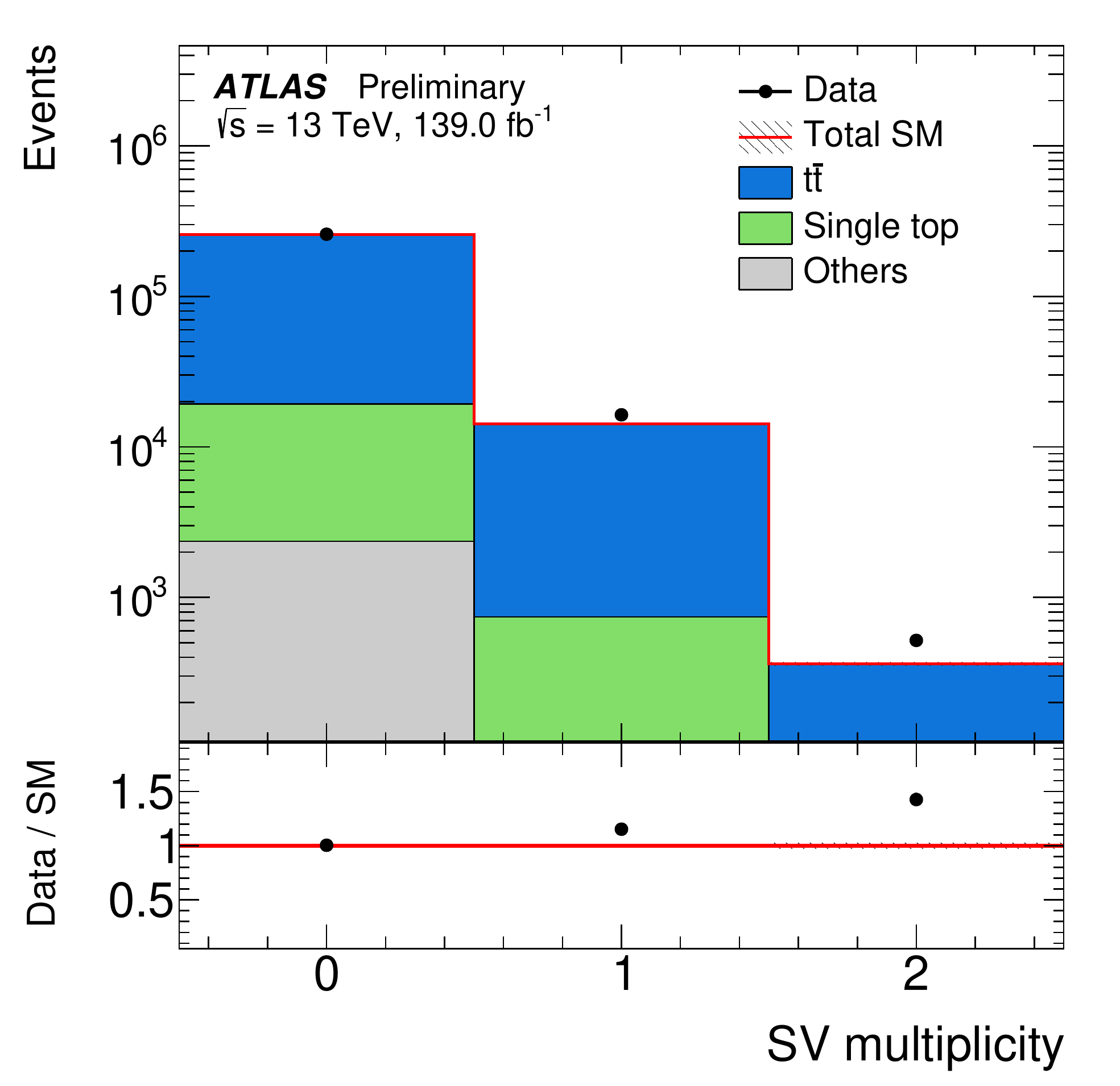}
\includegraphics[width=.40\textwidth]{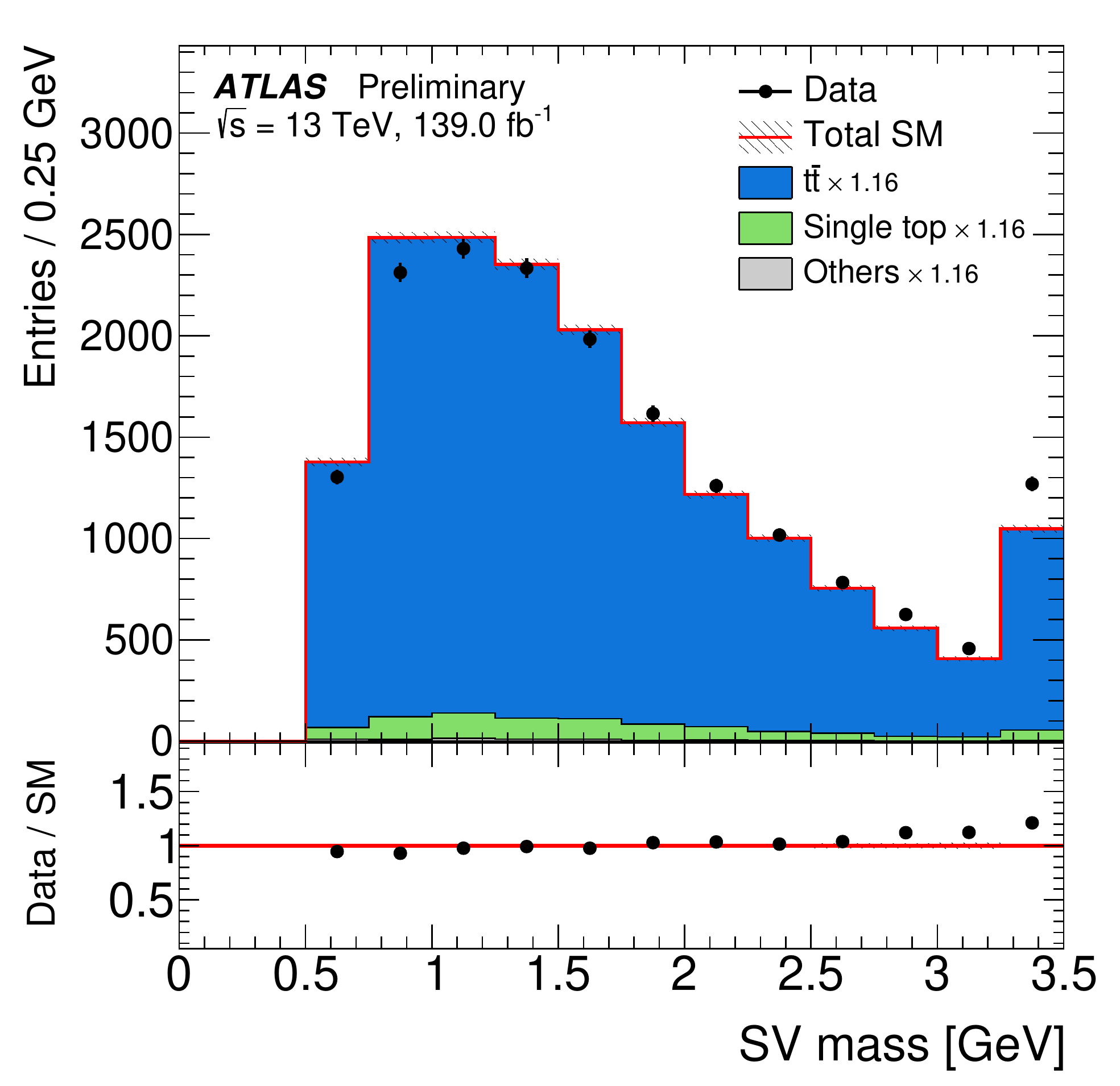}
\caption{Comparison between data and simulated events of (left)  the number of vertices per event and (right) the vertex mass for in a $t\bar{t}$ dominated region. In (right), the MC predictions have been scaled to match the overall yield observed in data. The "Others" category includes contributions from $V$+jets, diboson and $t\bar{t}+V$ production. Ratios between data and MC simulation are reported in the lower panel. Overflow events are included in the last bin~\cite{CONF:Softb2019}.}
\label{fig:softb_datamc}
\end{figure}

\noindent The modeling of secondary vertex kinematics is carefully validated comparing to observed data in a dileptonic $t\bar{t}$ enriched region. Events are required to contain a different flavor opposite sign lepton pair ($e\mu$), at least two calorimeter jets with $p_T$ $>$ 30 GeV, out of which one has to be $b$-tagged (MV2c10 with the 77\% efficiency working point). This yields a sample with a $t\bar{t}$ purity above 90\%. The normalization of simulated events do not match data perfectly, but the modeling is generally found to be very reasonable as shown in Figure~\ref{fig:softb_datamc}.

\section{Conclusion}
\noindent An update in the search for direct stop pair production in compressed-mass scenarios using 139 fb$^{-1}$ of $pp$ collisions at $\sqrt{s}=13$ TeV with the ATLAS detector was presented. For three-body stop decays, an RNN technique was employed to improve search sensitivity. The results significantly improved previous exclusion limits by excluding the stop mass region up to 720 GeV in the compressed three-body region under the assumption of 100\% {\it BR}($\tilde{t}_{1} \rightarrow bW\tilde{\chi}^{0}_{1}$). 
For four-body stop decays, a new $b$ identification algorithm was developed that targets $b$-hadrons in the $p_T$ range of 5 to 15 GeV and which is complementary to the standard ATLAS $b$-tagging algorithm. This new algorithm has identification efficiencies in the range 3-25\% and a misidentification rate of $8\times10^{-3}$ - $2.7\times 10^{-2}$. The identification of these soft $b$-hadrons is expected to greatly improve the stop search sensitivity in four-body decays.


\end{document}